\pdfoutput=1
\documentclass[12pt,a4paper]{article}
\usepackage{a4wide}

\usepackage{subcaption}
\usepackage[english]{babel}
\usepackage{amsmath}   
\usepackage{graphicx}  
\usepackage{bm}        
\usepackage{amssymb}   
\usepackage{epsfig}
\usepackage{epstopdf}
\usepackage{amsmath}
\usepackage{fontenc}
\usepackage[toc]{appendix}
\usepackage{color,soul} 
\usepackage{datetime}
\usepackage{physics}
\usepackage{enumerate}

\usepackage{footmisc}
\usepackage{textpos}
\usepackage{caption}
\usepackage{boxhandler}

\usepackage{hyperref}

\newcommand{\be}{\begin{equation}}
\newcommand{\ee}{\end{equation}}
\newcommand{\bea}{\begin{eqnarray}}
\newcommand{\eea}{\end{eqnarray}}
\newcommand{\bean}{\begin{eqnarray*}}
\newcommand{\eean}{\end{eqnarray*}}

\begin{document}



\vspace*{-14mm}

\begin{center}
\begin{center}
{\fontsize{20.74}{20.74} \bf{Fuzzballs and Observations}} \medskip \\
\end{center}
\vspace{2mm}

\centerline{{\bf Daniel R. Mayerson}}
\vspace{5mm}

\centerline{Institut de Physique Th\'eorique,}
\centerline{Universit\'e Paris Saclay, CEA, CNRS, }
\centerline{Orme des Merisiers,  F-91191 Gif sur Yvette, France}
\vspace{1mm}

%
{\footnotesize\upshape\ttfamily daniel.mayerson @ ipht.fr} \\

\vspace{3mm}
 
\textsc{Abstract}

\end{center}
 
\vspace{-3mm}
\noindent The advent of gravitational waves and black hole imaging has opened a new window into probing the horizon scale of black holes. An important question is whether string theory results for black holes can predict interesting and observable features that current and future experiments can probe.
In this article I review the budding and exciting research being done on understanding the possibilities of observing signals from fuzzballs, where black holes are replaced by string-theoretic horizon-scale microstructure.
In order to be accessible to both string theorists and black hole phenomenologists, I give a brief overview of the relevant observational experiments as well as the fuzzball paradigm in string theory and its explicitly constructable solutions called microstate geometries.

\vspace{-4mm}
 

\thispagestyle{empty}

\setcounter{tocdepth}{2}
\tableofcontents


\newpage
\section{Introduction}\label{sec:intro}

The last few years have given us an explosive expanse in the arsenal of tools we can use to observe and study black holes, with the advent of gravitational wave detections \cite{Abbott_2009} and black hole horizon imaging \cite{EHT2019a}. These observations have also opened the doors for gravitational phenomenology, where one can investigate if an astrophysical black hole shows departures from the general relativitistic expectations at the scale of its horizon.

One typically tries to model possible departures from general relativistic black hole physics by constructing and studying so-called \emph{exotic compact objects} or ECOs \cite{Cardoso:2017njb,Cardoso:2017cqb}; these are objects that are horizonless, but are compact enough to look like a black hole until approximately the scale of the would-be horizon. As such, their gravitational field will mimic that of the black hole, but still lead to significant, observable differences in experimental signals due to the non-trivial structure replacing the horizon.

An important parameter for any exotic compact object is the dimensionless \emph{closeness parameter} $\epsilon$ \cite{Cardoso:2019rvt}. If the would-be horizon sits at a scale $r_h$, then $\epsilon$ indicates the scale $r_h(1+\epsilon)$ at which the metric starts differing appreciably from the black hole metric.
There are many choices for how to precisely define $\epsilon$ \cite{Cardoso:2019rvt,Dimitrov:2020txx},\footnote{An alternative to $\epsilon$ is to define a length scale $R_a$ at which there are large modifications to the standard evolution. This is to be considered in conjunction with a second, ``hardness'' length scale $L$, which indicates the scale on which these new, modification effects vary \cite{Giddings:2014ova,Giddings:2016tla,Giddings:2019ujs}. Such ``hardness'' has been conjectured to be important in determining the energy that quanta must have in order to see effects from the non-trivial fuzzball structure \cite{Mathur:2020ely}.} although one certainly expects $\epsilon\ll 1$ for very compact ECOs\footnote{Note that Buchdahl's theorem in pure general relativity would imply $\epsilon \geq 1/8$ for static, spherically symmetric ECOs, when we take $r_h=2M$ to be the Schwarzschild radius \cite{PhysRev.116.1027,Cardoso:2019rvt}.} \cite{Cardoso:2019rvt}.

The actual ECO models that are typically studied in gravitational phenomenology are bottom-up models such as boson stars, gravastars, or wormholes;\footnote{Note that this refers to Solodukhin-type wormholes \cite{Solodukhin:2005qy} obtained by changing black hole metrics by hand, rather than the recent ones constructed in string theory such as in \cite{Gao:2016bin,Maldacena:2020sxe}.} for a more comprehensive list, see \cite{Cardoso:2019rvt}. These models are typically relatively simple to construct and study --- for example, a boson star is a self-gravitating soliton of a minimally coupled massive scalar field (possibly with self-interactions) \cite{PhysRev.172.1331,PhysRev.187.1767}. However, these bottom-up models all lack an interpretation or top-down derivation from a theory of quantum gravity such as string theory. They are also often plagued with fundamental problems for their actual physical viability. For example, a boson star has a maximal mass that is related to the inverse mass of the minimally coupled scalar field (taking for simplicity the case with no self-interactions). To be able to account for astrophysical black holes, this scalar must be ultralight with a mass below $\sim10^{-11}\,\text{eV}$ \cite{Cardoso:2019rvt}.

String theory provides a top-down, horizon-scale microstructure that replaces the black hole. In the so-called \emph{fuzzballs} \cite{Mathur:2005zp}, certain stringy mechanisms create topologically non-trivial bubbles threaded by fluxes.
This allows for the existence of structure that can be horizon-sized without collapsing under its own gravitational force \cite{Bena:2006kb,Bena:2017xbt,Gibbons:2013tqa,deLange:2015gca}. As such, these fuzzballs are an exciting target for gravitational phenomenology to investigate whether these top-down models of horizon-scale microstructure share observable properties of the bottom-up ECOs or lead to interesting, unexpected new phenomena.

Unfortunately, the fuzzballs that can be explicitly constructed (called \emph{microstate geometries}, see section \ref{sec:MGs}) are typically intricate solutions of supergravity theories with many fields excited, so it is often not trivial to investigate interesting observables therein. Nevertheless, the study of \emph{fuzzball observations} has taken flight recently, and the goal of this article is to review the recent progress of this field.

To be accessible for string theorists interested in making connections with observations, a brief introduction to the relevant observations is given in section \ref{sec:GWs} and \ref{sec:EHT}; the specific observables being discussed are also introduced briefly in their respective sections. For gravitational phenomenologists interested in understanding fuzzballs and their possible observational relevance, an introduction to the fuzzball paradigm and the explicitly constructable fuzzballs called microstate geometries is given in section \ref{sec:fuzzballs}. The rest of this paper reviews specific observables: echoes in section \ref{sec:echoesQNMs}; gravitational multipole moments and tidal Love numbers in section \ref{sec:multipolesTLNs}; and various aspects of geodesics in microstate geometries in section \ref{sec:geodesics}. Finally, an outlook for the future is briefly discussed in section \ref{sec:other}.



\subsection{Gravitational waves}\label{sec:GWs}

The first direct detection of gravitational waves was in September 2015 \cite{Abbott:2016blz} and heralded the start of a new era of astronomy where we are finally able to use the gravitational force to directly probe distant objects in the universe. The merger of two heavy objects such as black holes\footnote{Mergers of neutron stars also produce gravitational waves; this was first observed in 2017 \cite{LIGONS}. I will focus here on black hole (or ECO) mergers.}
produces a burst of gravitational waves which can be measured by detectors here on Earth; in particular, the current ground-based detectors, advanced-LIGO and VIRGO, operate at frequencies that will capture events coming from the merger of roughly similar size objects \cite{Abbott_2009}. The future space-based detector eLISA \cite{Danzmann_1996,Audley:2017drz} will also be receptive to signals coming from extreme mass ratio inspirals (EMRIs) \cite{Babak:2017tow}, where a solar-mass object orbits and falls into a much larger object such as a supermassive black hole.

Such gravitational wave signals arise from regions in spacetime where gravitational effects are large. As such, they are intimate probes of the near-horizon region of black holes, and deviations from the pure general relativistic predictions for gravitational wave signals can point to the existence of horizon-scale structure --- such as that coming from fuzzballs in string theory.

\begin{figure}[ht]\centering
 \includegraphics[width=0.7\textwidth]{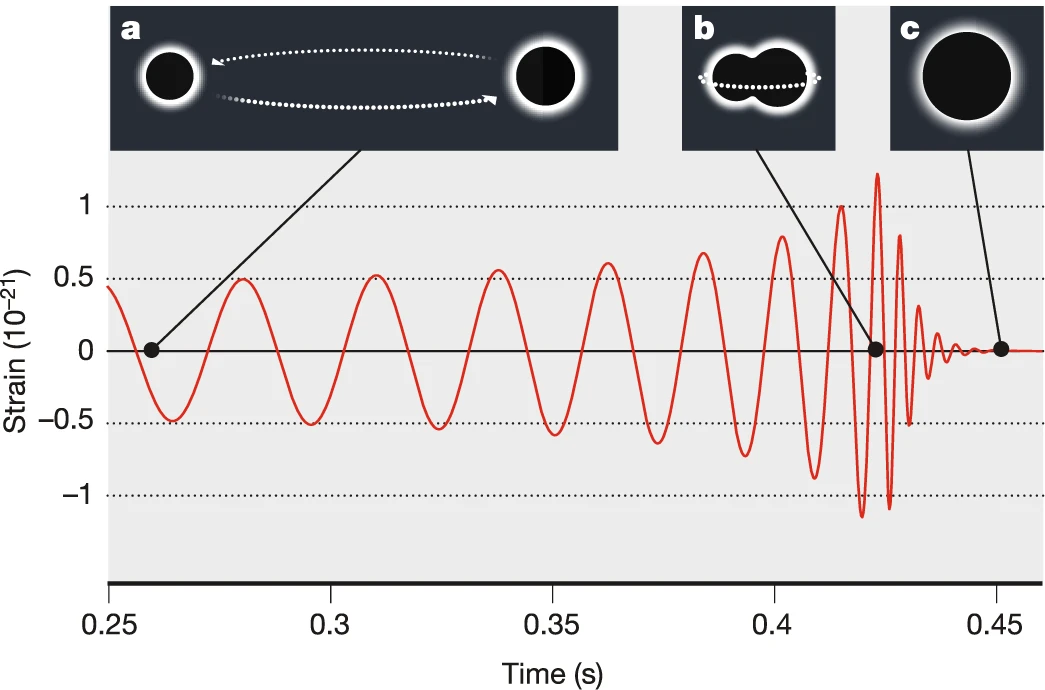}
 \caption[Gravitational wave signal]{Representation of the three phases of the merger of a binary system, and the gravitational waves emitted in the process. The three phases are (a) inspiral; (b) merger; (c) ringdown.\footnotemark
 }
 \label{fig:GWphases}
\end{figure}

The gravitational wave signal from a merger event can roughly be divided into three phases; see also fig. \ref{fig:GWphases}:
\begin{itemize}
 \item \textbf{Inspiral phase:} Here, the objects are still relatively far away from each other, and their interaction is still relatively weak. Perturbative general relativity methods can give reliable results. This phase is relatively long-lived; their mutual orbit decays slowly due to the emission of gravitational waves.
 \item \textbf{Merger phase:} The objects are now very close and eventually merge (or ``plunge'') into one, single object. There is no known way of getting an analytic handle on this phase; numerical relativity integration gives the only reliable results of this highly non-linear, strong field process.
 \item \textbf{Ringdown phase:} Once the final black hole has been formed, it relaxes to its final state in a relatively quick process. This can be well described by (decaying) perturbations on top of a single black hole geometry.
\end{itemize}
\footnotetext{This is fig. 1 of fig. \cite{GWdawn}, which is adapted from fig. 2 of \cite{Abbott:2016blz}, licensed under \href{https://creativecommons.org/licenses/by/3.0/}{CC BY 3.0}.}

The gravitational field is the strongest during the merger phase, so it is here that one would expect any departures from general relativity to be the most pronounced \cite{Giddings:2016tla}. Unfortunately, this is also the least accessible region to study --- in general relativity, this phase is only captured by complicated numerical simulations; an analytical understanding of how alternatives to black holes would behave in this phase seems beyond reach.

Thankfully, there are also potential signals in the other phases that are easier to study and model (semi-)analytically. For example, \emph{echoes} are a possible signal in the (late) ringdown phase; see section \ref{sec:echoesQNMs}. Other possibilities include signals coming from the multipole tidal deformations during the inspiral phase; see section \ref{sec:multipolesTLNs} (and especially section \ref{sec:TLNs}), and see also sections \ref{sec:tidalforces} and \ref{sec:other} for related topics.

A comprehensive overview and discussion of possible relevant black hole observables in gravitational wave signals can be found in \cite{Barack:2018yly}.
A review of the observational status of exotic compact objects (such as boson stars, gravastars, and wormholes) is found in \cite{Cardoso:2019rvt}.


\subsection{Black hole imaging}\label{sec:EHT}

In 2019, the first event-horizon-scale image of a black hole was made by the Event Horizon Telescope (EHT) \cite{EHT2019a}, a collaboration of many ground-based telescopes around the world combined into one, Earth-sized virtual telescope.

\begin{figure}[ht]\centering
\includegraphics[width=\textwidth]{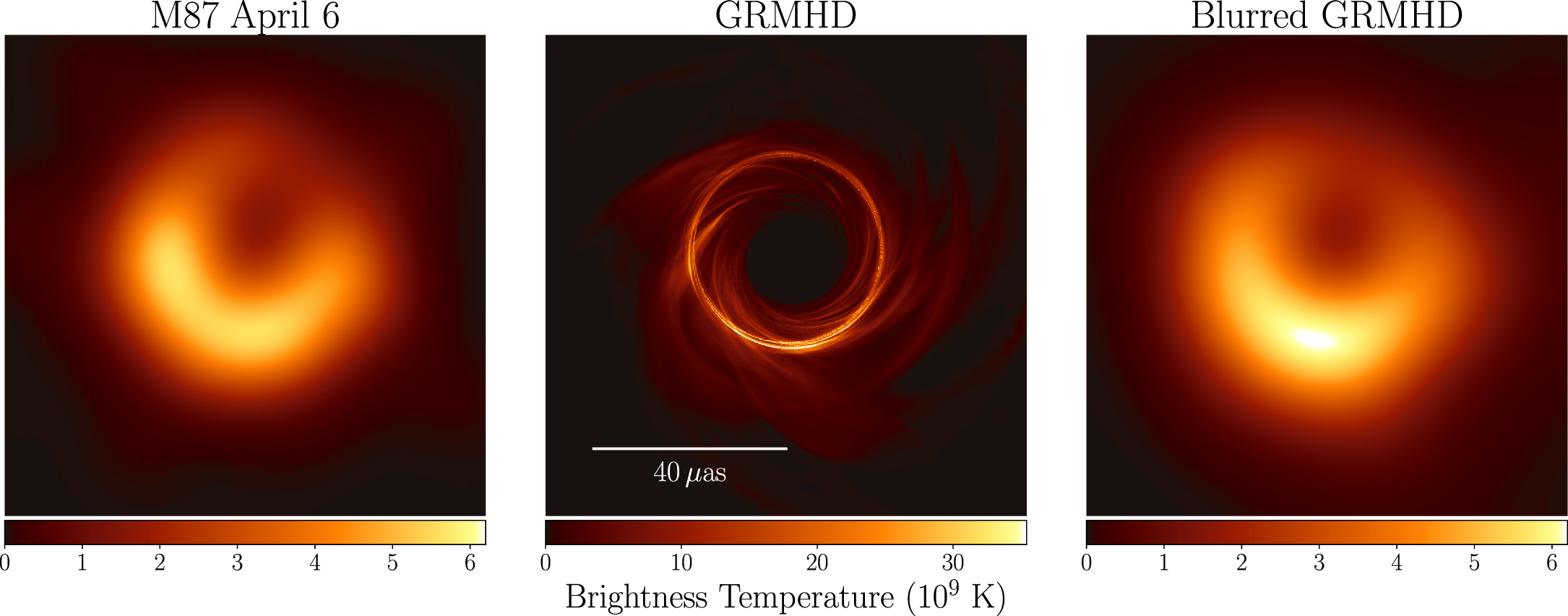}
\caption[EHT]{The black hole image as obtained by the EHT, and images from (one of) the best-fit simulations.
The actual observation is on the left; the image from the GRMHD simulation is in the middle; the image after blurring the GRMHD simulation image is on the right.\footnotemark}
\label{fig:EHT}
\end{figure}

Of course, one cannot directly see a black hole itself. Rather, the image of a black hole is sometimes more aptly termed its \emph{shadow},\footnote{\label{fn:shadow}I am being a bit loose with the terminology here. The actual \emph{shadow} of the black hole is usually defined to be the dark circular region in the center of the image of the black hole, which is actually not much influenced by the accretion disc \cite{Narayan:2019imo}. The outer edge of the shadow is called the \emph{photon ring}; for Schwarschild, this is $r_{\rm ph} =\sqrt{27} M$. The properties of this photon ring, if measured, can give very precise information on the black hole and any possible horizon-scale structure \cite{Narayan:2019imo,Gralla:2020srx,Gralla:2020yvo,Gralla:2020nwp,Johnson:2019ljv}.} and is the result of the complex interactions between the strong gravitational field of the black hole with the electromagnetic plasma interactions of the charged particles in its accretion disc \cite{EHT2019a,EHT2019e,Gralla_2019}. To model these effects, one must use intensive numerical general relativistic magnetohydrodynamic (GRMHD, also called gravitomagnetohydrodynamic) simulations of the accretion disc in the black hole background \cite{Porth2019}; then, (light) ray-tracing \cite{Gold2020} is applied to the system to extract the image of the black hole that can be seen from far away. The results of ray-tracing such a GRMHD simulation is the middle image of fig. \ref{fig:EHT}; after convolution with filters to simulate actual observation (due to telescope weather conditions, optical blurring, etc.), the end result is the right image of fig. \ref{fig:EHT}, which is a good match with the actual observation (left image of fig. \ref{fig:EHT}).

\footnotetext{This is fig. 1 from \cite{EHT2019e}, licensed under \href{https://creativecommons.org/licenses/by/3.0/}{CC BY 3.0}.} 

Note that these simulations consist of models with many tuneable parameters; comparison of the observed image with simulations does rule out many models \cite{EHT2019e,Psaltis:2020lvx}, but nevertheless still leaves much uncertainty in the selection of model and parameters. In fact, different models describing very different physical scenarios can be fit equally well to the observations \cite{EHT2019a,EHT2019e}.\footnote{The main uncertainty in model selection is not due to the underlying spacetime, of which it is pretty clear it must be (close to) the Kerr black hole spacetime. Rather, the details of the event horizon are effectively blurred by the complex physics of the accretion disc plasma \cite{Gralla:2019xty}. For example, the electrons can emit non-thermal radiation that is not accounted for in most simulations, leading to flares and hot spots \cite{Sironi:2014jfa,Ponti:2017grl,DoddsEden:2010wx,gravcol2018,Ripperda:2020bpz}.
This also makes it hard to detect or extract features of any possible horizon-scale microstructure replacing the black hole in the spacetime \cite{Gralla:2020pra}. Note that accretion disc plasma physics effects may also blur the ability of gravitational wave observations to perform precision strong-field gravity tests \cite{Cardoso:2020nst}.   (Many thanks to B. Ripperda for elaborating this issue to me.)} Nevertheless, it has been argued that deviations from general relativity can lead to observable differences in the observed EHT images. For example, quantum horizon-scale fluctuations, magnified by gravitational lensing effects, could lead to \emph{time-dependent} observable effects on the black hole shadow \cite{Giddings:2016btb,Giddings:2019jwy}.

The black hole image is very dependent on the behaviour of light rays or null geodesics in its geometry. To study and isolate lensing and shadow effects of a black hole or ECO metric without introducing the complicated plasma physics of an accretion disc, idealized images can be simulated using a ``screen'' of light. In such a setup, one considers the black hole (or ECO) in front of a bright plane (or screen) of uniform brightness, and without any further accretion disc surrounding the object. The resulting ray-traced imaged depends on the relative orientation of the plane, black hole, and observer at infinity \cite{Vazquez:2003zm,Amarilla:2015pgp}. Some analytic properties of such images can also be studied by analyzing the relevant geodesic equations \cite{Virbhadra:1999nm,Vazquez:2003zm}.

I will discuss various aspects of the behaviour of (null) geodesics in fuzzball geometries and how they may differ from their black hole counterparts in section \ref{sec:geodesics}.

\section{Fuzzballs \& Microstate Geometries}\label{sec:fuzzballs}
In this section, I will introduce the basic ideas behind fuzzballs and microstate geometries. The aim is to give enough context of these solutions to be able to follow the discussion in the next sections on fuzzball observables, but without inundating the reader with extraneous, unnecessary details of the often quite intricate solutions. First, I will set the stage in section \ref{sec:fuzzballpar}, giving the motivations and reasonings behind the fuzzball paradigm in string theory, and making the distinction between fuzzballs and microstate geometries more precise. The actual microstate geometries that we can study are introduced in section \ref{sec:MGs}. Finally, a few of the most important pitfalls and limitations of microstate geometries are mentioned in section \ref{sec:limitations}.

\subsection{The fuzzball paradigm}\label{sec:fuzzballpar}
Black holes have been familiar objects in physics for over a century. Their properties in general relativity have been studied in great detail, leading to many deep insights into their nature such as the no-hair theorems in four dimensions \cite{nohair1,nohair2,nohair3,Mazur:2000pn}. Even semi-classical properties of black holes are well known, such as Hawking's famous calculation showing black holes emit thermal radiation \cite{Hawking:1974sw}.

Nevertheless, the existence and properties of black holes lead to fundamental, deep problems in our understanding of their physics and the behaviour of quantum gravity in general. The most striking of these problems is called \emph{the information paradox}. This is perhaps most succintly explained as a two-pronged problem (following \cite{Mathur:2017fnw}): 
\begin{itemize}
 \item Classically, all information is necessarily trapped inside the black hole after it falls in. Thus, the information cannot escape as the black hole evaporates through Hawking radiation. How can this information escape anyway, and/or what happens with the information stored inside the black hole after it has evaporated away completely?
 \item The process of Hawking radiation creates entangled pairs of particles at the horizon; of each pair, one escapes outside the black hole, leading to the thermal spectrum of radiation observed outside the black hole.
 At the same time, the entanglement between the black hole and its radiation increases monotonically in time. What happens to this large entanglement after the black hole has evaporated?
\end{itemize}
An obvious critique of these problems is that Hawking's calculation is a leading order, semi-classical calculation which ignores backreaction of the radiation on the background metric. In particular, one may hope that small, quantum corrections of horizon-scale quantum physics might be able to accumulate over time, resolving the paradox by somehow leading to all of the information and entanglement being able to escape from the black hole. However, a theorem by Mathur proves precisely that such \emph{small corrections at the horizon are not enough} \cite{Mathur:2009hf}, so that this does not give a way out.

There are roughly three ways that this paradox may be resolved:
\begin{itemize}
 \item The existence of \emph{remnants}: there is no escape of information or entanglement from the black hole. Rather, when the object reaches a certain small Planck-scale size, quantum effects stop the evaporation; the information is forever trapped in the resulting tiny object. This scenario is rather unappetizing as it leads to a necessarily infinite degeneracy of states for Planck-scale energies, and moreover can be seen not to be allowed in string theory using AdS/CFT \cite{Mathur:2017fnw}. A so-called ``firewall'' \cite{Almheiri:2012rt,Mathur:2013gua} could be interpreted as a kind of remnant.
 \item Allowing \emph{non-local effects}: Mathur's theorem is valid for any small and \emph{local} quantum corrections at the horizon, but a \emph{non-local effect} is not ruled out by this. Indeed, Papadodimas and Raju argue that long-range, non-local entanglement and interactions can be found when carefully considering modes near the horizon in AdS/CFT \cite{Papadodimas:2012aq,Papadodimas:2013jku,Papadodimas:2013wnh,Papadodimas:2015jra}; these nevertheless do not lead to violations of causality or the equivalence principle (under the form of black hole complementarity) for semi-classical observers. Recently, a similar \emph{island} proposal has been made \cite{Almheiri:2019psf,Almheiri:2019hni,Penington:2019npb,Almheiri:2019yqk}, where the particles and information of the radiation \emph{outside} the black hole, and the particles and information \emph{inside} the black hole are effectively identified as one and the same. Another approach using non-local physics to describe quantum black holes is the ``non-violent non-locality'' advocated in \cite{Giddings:2006sj,Giddings:2012bm,Giddings:2012gc,Giddings:2013kcj,Giddings:2016plq}. I will not discuss these ideas here further, except to note that they are not (all) necessarily mutually exclusive with the fuzzball paradigm (for example, see \cite{Guo:2017jmi}).
 \item \emph{New physics at the horizon scale}, in other words, the physics at the horizon is altered significantly compared to our expectation from black holes in general relativity. In a nutshell, this is the basic ingredient of the \emph{fuzzball paradigm}.
\end{itemize}
In the fuzzball paradigm, then, the black hole with its horizon is entirely replaced by a fuzzball, which is a horizonless black hole microstate that has non-trivial structure which differs from the black hole geometry at the would-be horizon scale. Of course, as we move further away from the horizon scale, the effects of the microstructure should become less and less visible, meaning that the black hole and its fuzzball microstates share the same asymptotic structure --- they look the same from far away.

Introducing large corrections at the event horizon scale seems counterintuitive at first thought. One expects quantum corrections to arise in a black hole geometry where the curvature is large, which may be a small region around the singularity at the center of a black hole, but certainly not at the outer horizon\footnote{One \emph{would} perhaps expect quantum corrections at \emph{inner} horizons, which are Cauchy horizons and may be subject to quantum instabilities \cite{Hollands:2019whz,penrosecollapse}.} when the black hole is of a reasonable size. However, another crucial ingredient that we must not forget is the \emph{size of the phase space} that is available to matter at the horizon scale. This is best illustrated with the example of a collapsing shell of dust; once it has reached the scale of its would-be horizon, the shell has an enormous number of fuzzball states available to it that it can quantum tunnel into. Even though the tunneling amplitude to tunnel into any given single state is incredibly small, the enormous number of states that are available add up to an $\mathcal{O}(1)$ probability that the collapsing shell \emph{will tunnel} into one of these states \cite{Kraus:2015zda,Mathur:2008kg,Mathur:2009zs}.


\subsection{Microstate geometries}\label{sec:MGs}
I have introduced fuzzballs rather broadly and vaguely as microstates which have \emph{horizon-scale microstructure}, meaning they look like the black hole they are meant to replace from far away, but have large differences at the would-be horizon scale. The most general microstate or fuzzball for an arbitrary black hole will undoubtably be a very complicated, quantum stringy object in string theory for which we do not (yet) have the tools to describe. However, in certain cases, and for certain black holes, we are able to explicitly construct fuzzballs that are well-behaved solutions in a supergravity theory which is obtained as a low-energy effective theory of string theory. These solutions are called \emph{microstate geometries}, and are our main tools to study the fuzzball paradigm explicitly in string theory, including its possible observable effects.

Following \cite{Bena:2013dka}, a microstate geometry is a solution of a supergravity theory with the following properties:
\begin{itemize}
 \item It is \emph{horizonless}.
 \item It carries the same charges (electromagnetic as well as angular momentum) as a given black hole.
 \item It is smooth, meaning the geometry has no singularities. Relaxing this condition is sometimes possible, although one could then qualify such singular solutions \emph{microstate solutions} \cite{Bena:2013dka}. For example, some smooth microstate geometries in five dimensions also have an interpretation in four dimensions, where they have certain well-understood singularities.
 \item It is valid within the supergravity regime. This is related to the previous point, but can also be more subtle \cite{Chen:2014loa,Raju:2018xue}; essentially, one must be careful not to trust scales in the geometry which are below the Planck or string scale and thus are susceptible to receiving (large) quantum or stringy corrections.
\end{itemize}
Hence, microstate geometries are precisely the fuzzballs in string theory that we are able to explicitly construct and analyze using supergravity and string theory methods. The microstate geometries that I will consider in this paper can be divided into two main classes, which can respectively be called \emph{multicentered bubbled} geometries and \emph{superstrata} geometries.

\subsubsection{Multicentered bubbled geometries}\label{sec:multicenter}
The multicentered bubbles geometries are smooth, horizonless solutions in five-dimensional supergravity \cite{Bena:2007kg}. They can have either $\mathbb{R}^{4,1}$ or $\mathbb{R}^{3,1}\times (S^1)_\psi$ asymptotics. In the former case, they can be seen as microstates of a five-dimensional, three-charge supersymmetric BMPV black hole \cite{Breckenridge:1996is}. In the latter case, they can be reduced down over the $\psi$ circle to four dimensions and are microstates of a supersymmetric, static four-dimensional black hole. These four-dimensional solutions are of the type which were first constructed by Denef and Bates \cite{Bates:2003vx}.

These multicentered bubbled geometries are characterized by a number of ``centers'', which are points (in a $\mathbb{R}^3$ subspace of the geometry) where the $(S^1)_\psi$ circle pinches off in a way that leaves the spacetime completely smooth.\footnote{Not all centers are smooth in five dimensions; the charges or parameters of a center must obey certain conditions in order to avoid being singular \cite{Bena:2007kg}.} Between each pair of such centers, there is a ``bubble'', a topological $S^2$ that is prevented from collapsing by the presence of magnetic flux through it. In this way, the presence of extra dimensions allows (through the formation of this topology) for non-trivial, massive structure to exist at macroscopic scales without collapsing. Note that when the asymptotic $\mathbb{R}^{3,1}\times (S^1)_\psi$ geometries are reduced (over the $(S^1)_\psi$ circle) to four-dimensions, the centers become (naked) singularities.

The positions of the centers are not arbitrary; they must satisfy coupled, non-linear \emph{bubble equations} involving each center's charges as well as the intercenter distances. The most interesting multicentered solutions are called \emph{scaling solutions}: this is a continuous family of solutions for a parameter $\lambda$, including a so-called \emph{scaling limit} or \emph{scaling point} $\lambda\rightarrow 0$ at which the intercenter distance vanishes and thus the centers coincide. It is precisely in this limit that the microstate approaches the corresponding black hole geometry.

\begin{figure}[ht]\centering
 \includegraphics[width=0.8\textwidth]{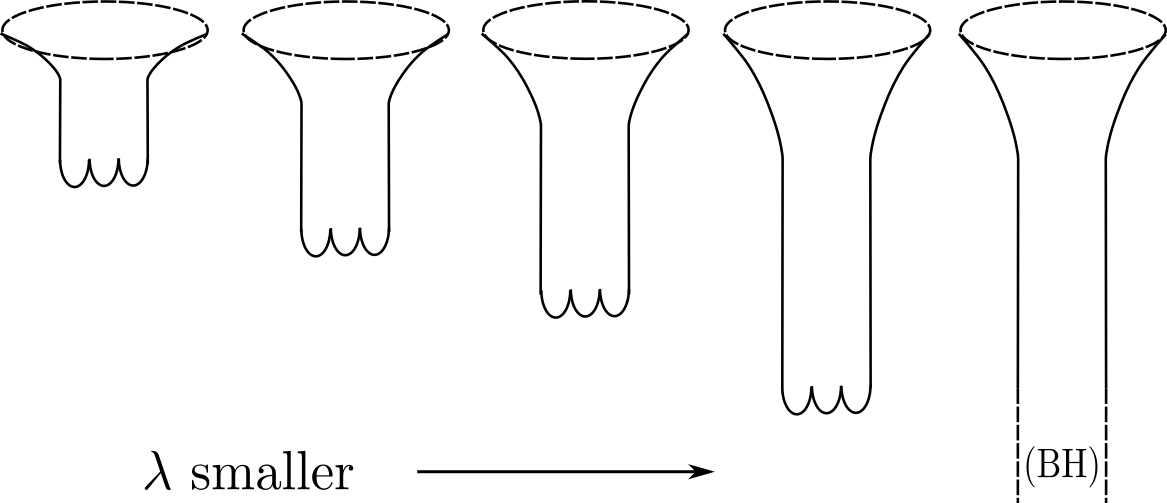}
 \caption{A representation of how the multicentered bubbled microstate geometry's redshift throat becomes deeper as the scaling limit $\lambda\rightarrow 0$ is approached; the corresponding black hole on the far right has an infinite throat.
 }
 \label{fig:scalinglimthroat}
\end{figure}

The black hole geometry, being extremal, has an infinite redshift $AdS_2$ ``throat'' at its horizon.
As we approach the scaling point in a multicentered geometry, the bubbles sit at the bottom of a deeper and deeper redshift throat (a deeper throat means the difference in redshift factors inside and outside the throat is larger). In the scaling limit $\lambda\rightarrow 0$, this approaches the infinite redshift throat of the black hole horizon. See fig. \ref{fig:scalinglimthroat} for a schematic depiction of this scenario.

A comprehensive review of multicentered bubbled geometries, their construction and properties is given from a five-dimensional perspective in \cite{Bena:2007kg}; see especially sections 4.1, 5.1, 6.2-6.4 for the construction of the solutions. Alternatively, section 4.1 of \cite{Bena:2020uup}, for example, contains all of the information needed for the construction of (only) the metric in a four-dimensional perspective.

\subsubsection{Superstrata}\label{sec:SS}
The five-dimensional version of the multicentered bubbling geometries introduced above correspond to microstates of the three-charge, rotating supersymmetric black hole in five dimensions called the BMPV black hole. However, it can be shown that these bubbling geometries only correspond to a small fraction of the total microstates of this black hole \cite{Bena:2010gg}. The microstate geometries called \emph{superstrata} corrrespond to a much larger family of microstates of this same black hole. Heuristically, one can picture a superstratum geometry as having a \emph{single} topological $S^2$ bubble, similar to a two-centered geometry of the multicentered geometries described above. However, in contrast to the multicentered geometries, the charge density on this single bubble has a non-trivial oscillating profile. It turns out that this single oscillating bubble gives rise to a much richer and larger phase space of solutions than the multiple (rigidly charged) bubbles of the multicentered geometries described above \cite{Bena:2014qxa,Shigemori:2019orj}. However, note that even these superstrata solutions are still not nearly enough to account for the BMPV entropy \cite{Strominger:1996sh,Shigemori:2019orj,Mayerson:2020acj}.

\begin{figure}[ht]\centering
  \includegraphics[width=0.7\textwidth]{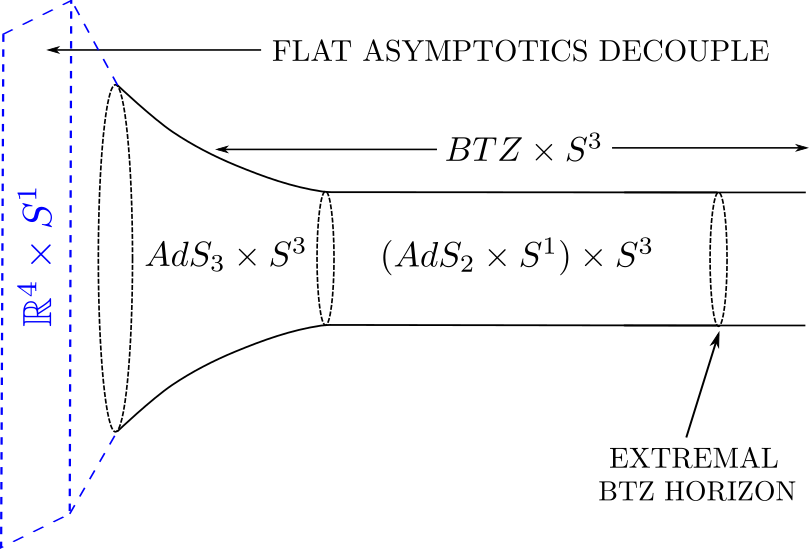}
 \caption{In the decoupling limit of the BMPV black hole, the BTZ black hole with $AdS_3$ asymptotics (and an $AdS_2\times S^1$ near-horizon redshift throat) is decoupled from the asymptotically flat space.}
 \label{fig:decouplingBTZ}
\end{figure}

The superstrata geometries are solutions of \emph{six}-dimensional supergravity; the extra (sixth) dimension serves to support non-trivial momentum wave profiles along it (i.e. the charge density ``fluctuations'' on the bubble). The corresponding five-dimensional black hole, the BMPV black hole, becomes a six-dimensional black string which wraps the compact $y$ circle and has asymptotic topology $\mathbb{R}^{4,1}\times (S^1)_y$, where the $y$ circle has radius $R_y$.

\begin{figure}[ht]\centering
 \includegraphics[width=0.7\textwidth]{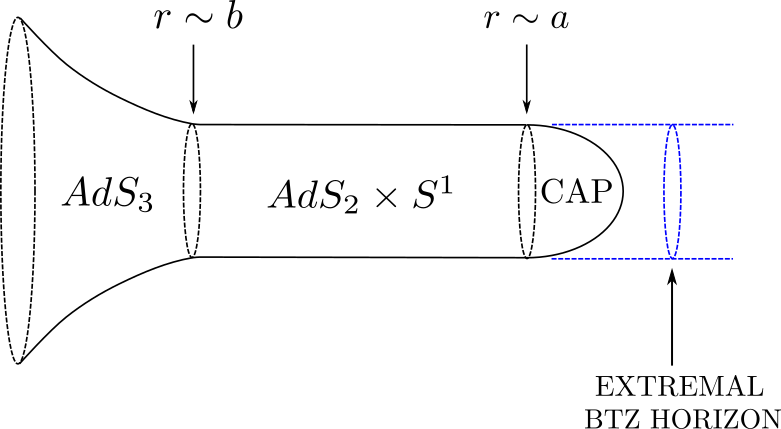}
 \caption{A superstratum geometry with $AdS_3$ asymptotics. At $r\sim b$, a BTZ-like $AdS_2\times S^1$ redshift throat appears. The throat ends at $r\sim a$ with a smooth cap; the total length of this throat as measured by the difference in redshift is $\sim b^2/a^2$. By contrast, the BTZ black hole has an infinite redshift at the horizon and thus an infinite length throat.
 }
 \label{fig:10nsuperstrata}
\end{figure}

Both the BMPV black string and the superstrata have a so-called decoupling limit where the asymptotically flat geometry is decoupled from a region which is asymptotically $AdS_3\times S^3$. For the black hole, this corresponds to the standard holographic decoupling limit \cite{Maldacena:1997re} where the resulting $AdS_3$ geometry is the extremal BTZ black hole with an infinite $AdS_2\times S^1$ redshift throat; see fig. \ref{fig:decouplingBTZ}. For a superstratum, the resulting $AdS_3$ geometry looks like the BTZ black hole geometry asymptotically, but  after a (long but not infinite) $AdS_2\times S^1$ throat, the superstratum ends in a smooth ``cap'' geometry; see fig. \ref{fig:10nsuperstrata}. The $AdS_3$ decoupling limit allows for a holographic interpretation of the superstrata in the dual $CFT_2$; the map between holography and CFT states is very well understood for superstrata \cite{Kanitscheider:2006zf,Kanitscheider:2007wq,Giusto:2013bda,Giusto:2015dfa,Shigemori:2019orj}. By contrast, the holographic dual interpretation of solutions with more than two centers is not known.

It is much easier to work with (and construct) superstrata geometries with $AdS_3\times S^3$ asymptotics; in fact, not all known asymptotically $AdS_3\times S^3$ superstrata have known asymptotic flat counterparts since they are technically much more difficult to construct. Thus, much of the study of superstrata has been using the asymptotically $AdS_3\times S^3$ solutions, which should be seen as studying the physics of the region (very) near the would-be horizon.

Superstrata geometries are typically labelled by their \emph{mode numbers}, which roughly can be seen as the modes in the Fourier decomposition of the charge density profile on the single $S^2$ bubble. There are two families of complex three-dimensional mode numbers: $b_{k,m,n}$ and $c_{k,m,n}$, with $k>0, m\geq 1, n\geq 1$ and either $k\geq m$ (for $b_{k,m,n}$) or $k>m$ (for $c_{k,m,n}$). The $b_{k,m,n}$ coefficients are referred to as the ``original'' $(k,m,n)$ superstrata  (or just superstrata) whereas the $c_{k,m,n}$ are the ``supercharged'' superstrata.

The most general superstrata would be a geometry corresponding to an arbitrary superposition of any number of original $(k,m,n)$ as well as supercharged $(k,m,n)$ superstrata, but the construction of such a geometry is beyond current techniques. The arbitrary \emph{single-mode} (either original or supercharged) for one single $(k,m,n)$ superstratum geometry can be constructed \cite{Bena:2015bea}. In principle, it is known how to construct the most general two-mode superposition geometry \cite{Heidmann:2019zws}. Certain multi-mode superstrata geometries are also explicitly known \cite{Heidmann:2019xrd,Mayerson:2020tcl}; the simplest of these is the arbitrary superposition of $(1,0,n)$ superstrata with different $n$. This superposition can be packaged into an arbitrary holomorphic function of one complex variable. It is thought that the most general superstratum will correspond to a geometry depending on two holomorphic functions of three complex variables, one for the original modes and one for the supercharged modes \cite{Heidmann:2019zws,Heidmann:2019xrd}.

The most important parameters in a superstratum solution are the D1- and D5-brane charges $Q_1$ and $Q_5$ and two parameters $a$ and $b$, where $b$ is obtained from the mode numbers discussed above:
\be b^2\sim \sum_{k,m,n}\left(\left|b_{k,m,n}\right|^2 + \left|c_{k,m,n}\right|^2\right), \ee
and where the parameters much satisfy the constraint:
\be \frac{Q_1 Q_5}{R_y^2} = a^2 + \frac12 b^2.\ee
Roughly speaking, $a$ controls the left-moving five-dimensional angular momentum $J\sim a^2$ and $b$ is related to the momentum charge $Q_P\sim b^2$.
The superstrata which most closely approximate the BMPV black hole are the ``deep, scaling'' geometries with $a^2\ll b^2$ \cite{Bena:2017xbt,Bena:2016ypk}; for these, we have the approximate relations \cite{Tyukov:2017uig}:
\be b^2\approx \frac{2Q_1 Q_5}{R_y^2}, \qquad \frac{a^2}{b^2} \approx \frac{J}{N_1N_5},\ee
where $N_1,N_5$ is the quantized number of D1,D5 branes.\footnote{The quantized number of D1,D5,P excitations is related to the supergravity charges as $Q_1 = ((2\pi)^4 g_s \alpha'^3/V_4)\, N_1$, $Q_5 = (g_s \alpha')\, N_5$, and $Q_P = (\alpha'^4/(V_4 R_y^2))\, N_P$, where $g_s$ is the (ten-dimensional) string coupling, $\alpha'=l_s^2$ is the string length squared, and $V_4$ is the volume of the compact four-dimensional manifold ($T^4$ or $K3$) that the D5-branes wrap.} The parameters $a,b$ determine the length of the $AdS_2$ throat \cite{Bena:2018bbd}; this can be most conveniently expressed as the difference in redshift factors between the top and bottom which scales as $\sim b^2/a^2$; see also fig. \ref{fig:10nsuperstrata}.

For a review and current status of superstrata geometries, see \cite{Shigemori:2020yuo}; see especially section 4.3 therein for the explicit supergravity solutions.

\subsubsection{Other microstate geometries}\label{sec:otherMG}

The multicentered bubbled geometries and the superstrata are by far the largest families of microstate geometries that have been constructed. However, all of these suffer the main drawback that they are supersymmetric and thus are also microstates of a supersymmetric black hole (either a static, four-dimensional black hole, or the five-dimensional BMPV black hole).

It is very difficult to construct microstate geometries that are non-supersymmetric, since then typically the full equations of motion in supergravity must be solved (as opposed to the simpler, first-order supersymmetry equations). Nevertheless, certain non-supersymmetric microstate geometries have been constructed \cite{Jejjala:2005yu,Bena:2009en,Bena:2009qv,DallAgata:2010srl,Vasilakis:2011ki,Bena:2011fc,Bena:2012zi,Mathur:2013nja,Banerjee:2014hza,Bena:2015drs,Bena:2016dbw,Bossard:2017vii}, although they are usually \emph{overspinning}, meaning there exists no black hole with the same charges as it would have too much angular momentum to have a well-defined horizon area (similar to $a>M$ for a Kerr black hole).

One important family of non-supersymmetric microstates to mention are the \emph{almost-BPS multicentered microstates} \cite{Bena:2009en}. These microstates are constructed by solving a clever modification of the supersymmetry equations. In essence, these microstate solutions \emph{locally} solve the supersymmetry equations but are not \emph{globally} supersymmetric. As their name suggests, these microstate geometries are non-supersymmetric siblings of the multicentered bubbled geometries discussed above, and share many of their properties. In four dimensions (with $\mathbb{R}^{3,1}\times (S^1)_\psi$ asymptotic topology as solutions in five-dimensional supergravity), these microstate geometries can be seen as microstates of an \emph{almost-BPS black hole} \cite{Bena:2009en,Goldstein:2008fq,Bena:2009ev}, an extremal but non-supersymmetric black hole in four dimensions which also has a non-zero angular momentum (as opposed to the necessarily static four-dimensional supersymmetric black hole).

These almost-BPS solutions can be seen to be ``seed'' solutions that generate many more non-supersymmetric solutions through certain $U$-duality transformations and so-called generalized spectral flows \cite{Bena:2012wc,Bena:2008wt}. In this way, microstate geometries can be constructed with the asymptotics of near-horizon extremal (five-dimensional) Kerr (NHEK), effectively providing the first indirect examples of microstate geometries for Kerr black holes \cite{Heidmann:2018mtx}. However, it is still an open problem to understand how one can ``glue'' the NHEK asymptotics of these NHEK microstate geometries back to flat space to obtain microstate geometries of the asymptotically flat Kerr black hole geometry.


\subsection{Limitations}\label{sec:limitations}
The most important limitation of the current microstate geometry program was already mentioned above: \emph{no microstate geometries for realistic (non-supersymmetric and non-extremal) black holes are currently known}. Of course, this makes it challenging to compare any observational results or signals that one might find in microstate geometries to actual observations of astrophysical black holes. 

Nevertheless, at the very least serving as toy models, the (known) microstate geometries can give qualitative, general insights into the physics of black holes and the microstructure that we expect to see at their horizons. This is similar to the role of holography for heavy-ion collisions, where AdS/CFT provided the only analytic insights available into the smallness of the viscosity over entropy ratio in the strongly coupled quark-gluon plasma \cite{Policastro:2001yc,Son:2007vk,Cremonini:2011iq,Shuryak:2004cy}. Microstate geometries will point the way to measurable observables and provide estimates for possible observable deviations from general relativity.

\subsubsection{Formation \& evolution}
When a solution is supersymmetric, such as the main families of microstate geometries described above, it is effectively non-dynamical. This means there is no strict sense in which questions regarding their formation or evolution can be framed or analyzed. However, by introducing a non-supersymmetric probe in such backgrounds, one can raise the energy slightly and view the probe dynamics as a prototype for the evolution of a near-extremal state consisting of geometry and probe together. In this way, analytic calculations of the formation rate for multicentered microstate geometries were performed \cite{Bena:2015dpt}; these explicit calculations were then used as input in a more elaborate setup using networks to study in detail the possible formation, evolution, and subsequent relaxation to a steady state of such microstate geometries \cite{Charles:2018oob}. A different approach to studying properties of merging black holes and their microstates in string theory is to consider collisions of stringy objects in the string worldsheet theory. This approach was used to study stringy corrections to supersymmetric black hole mergers using vertex operators \cite{Addazi:2020obs}, and to study high-energy string-string collisions \cite{Addazi:2016ksu}.

\subsubsection{Typicality}\label{sec:MGtypicality}
There is only one system in which the known microstate geometries are ``enough'' to account for the full entropy: the D1/D5 system with their corresponding smooth Lunin-Mathur supertube microstate geometries \cite{Lunin:2001jy,Lunin:2002iz,Rychkov:2005ji,Kanitscheider:2006zf,Kanitscheider:2007wq,Krishnan:2015vha}.\footnote{Even here, the Lunin-Mathur geometries with the most typical supertube profile are of a stringy size and so fall outside of the regime where the supergravity approximation is valid. Moreover, the most typical states will be quantum superpositions of the states described by these Lunin-Mathur supertube geometries and so are not themselves described by a geometry \cite{Lunin:2002qf,Mathur:2005zp,Chen:2014loa,Raju:2018xue}. Additionally, the D1/D5 system does not correspond to a black hole of finite size; rather, the ensemble of D1/D5 states corresponds to a singular geometry where the would-be horizon has zero size.} For all other microstate geometries, it is clear that the currently known families contain a parametrically smaller number of states than the total number of states suggested by the corresponding black hole entropies \cite{deBoer:2009un}.

This raises the important question: are the microstate geometries \emph{typical} microstates, in the sense that they represent and share properties with most of the microstates of the black hole \cite{Balasubramanian:2007qv,Bao:2017guc,Michel:2018yta,Guo:2018djz}? Or are they \emph{atypical}, having very different properties than most of the other black hole microstates? In the latter case, it may not make as much sense to use microstate geometries as a prototype for what we could expect to realistically observe when we look at real black holes, since they would then correspond to an atypical, almost-negligible portion of the phase space.

It has been argued that the known microstate geometries are atypical \cite{Raju:2018xue}, which may cast doubt on whether such microstate geometries can share meaningful properties with typical fuzzballs that one would expect to observe in nature. On the other hand, there are also arguments and examples that suggest that microstate geometries can share at least some properties of typical microstates; perhaps most notably, the energy gap in deep microstate geometries was shown to be typical \cite{Bena:2007qc,Bena:2006kb,deBoer:2008zn,Tyukov:2017uig}.
Here, I would like to advocate taking an optimistic viewpoint: even though many microstate geometries may not be typical in some ways, they still serve as useful tools to identify universal properties of black hole microstates \cite{Dimitrov:2020txx}, and can qualitatively point the way to interesting possible observable quantities in current and future observations which may indicate the existence of horizon-scale microstructure \cite{Hertog:2017vod}.

\section{Echoes \& Quasinormal Modes}\label{sec:echoesQNMs}

The ringdown phase of a black hole merger (see section \ref{sec:GWs} and fig. \ref{fig:GWphases}) is governed by the physics of the final object, and in particular its light ring \cite{Cardoso:2016rao,Cardoso:2016oxy}, which is the smallest radius at which photons can be in an (unstable) orbit.\footnote{Unfortunately, there is some overloading of the terms ``light ring'' or ``photon ring'' in the literature. The conventions followed by the EHT collaboration \cite{EHT2019a,EHT2019e,Narayan:2019imo} are that the photon ring is the edge of the black hole shadow --- for Schwarschild, this is $r=\sqrt{27}M$; see also footnote \ref{fn:shadow}. However, in most gravitational wave papers such as \cite{Cardoso:2016rao}, the convention is that the light ring is the smallest photon orbit radius, which for Schwarzschild is $r=3M$.} If the final object is a black hole, the ringdown is simply a fast relaxation to the final, stationary black hole state, after which there is no more signal --- everything has either fallen into the black hole, or radiated away.

\begin{figure}[ht]\centering
 \includegraphics[width=0.8\textwidth]{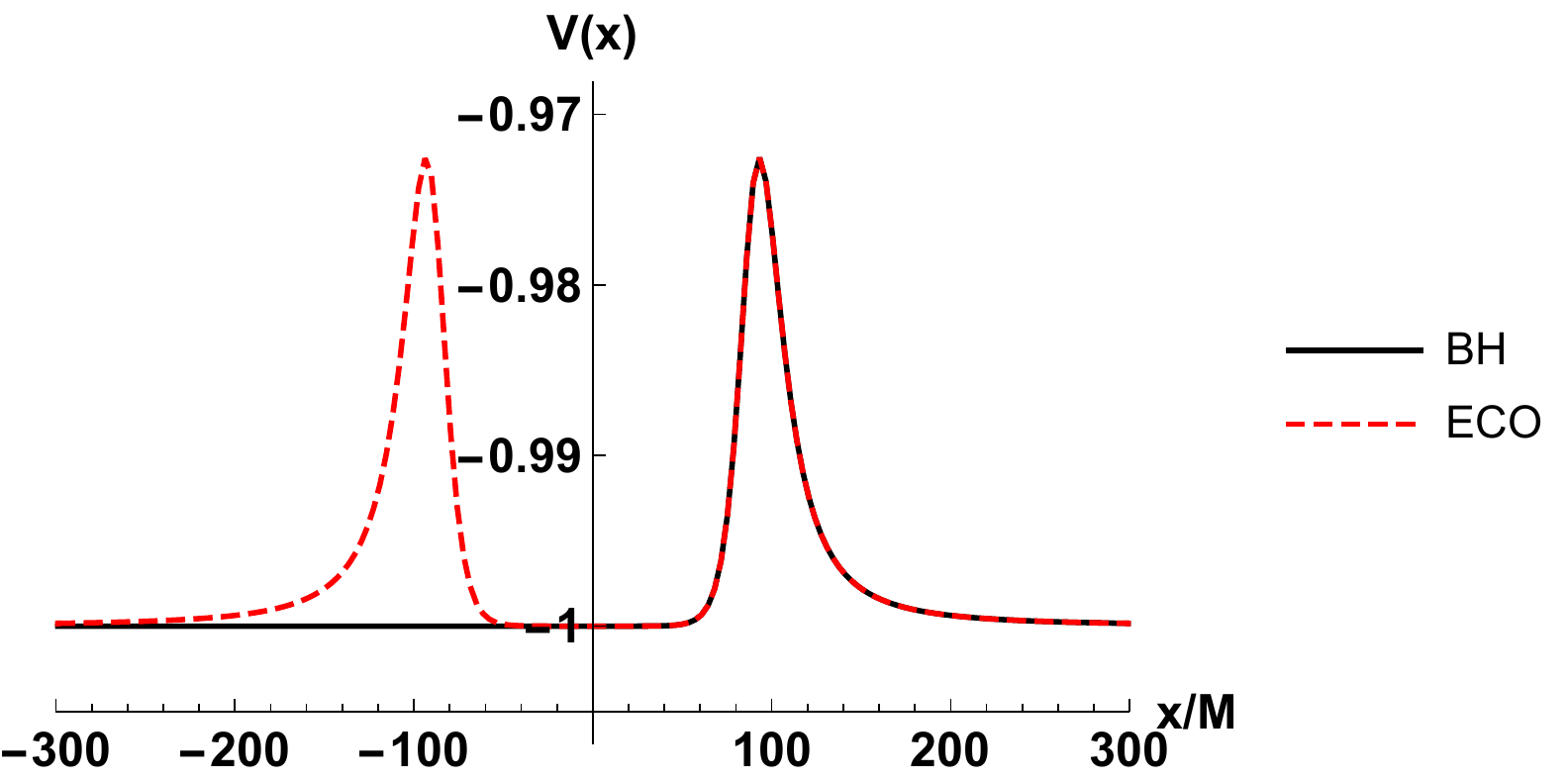}
 \caption[Effective potential]{The effective potential that a wave sees in a (Schwarschild) black hole background (solid black line), and in a particular exotic compact, horizonless object or ECO\footnotemark (dotted red line). Here, $x$ is an appropriate tortoise coordinate such that $\lim_{r\rightarrow\infty} V(x(r)) = -1$. The potential that the scalar feels in the ECO background mimicks that of the black hole background up until the horizon-scale structure at $x\sim -100M$. This structure leads to an effective cavity size of $L\sim  186M$.
 }
 \label{fig:echopotential}
\end{figure}

However, when we replace the black hole by a compact, horizonless object (ECO), the effective potential that a wave perturbation sees in the neighborhood of the would-be horizon is no longer simply flat. Instead, in several ECO models there is an effective potential barrier ``inside'' the object that will reflect any incident wave back out --- see fig. \ref{fig:echopotential}. This means that a wave packet will get effectively stuck inside this potential well for a long time, reflecting back and forth between the barriers. Every time the wave reflects against the ``outer'' barrier, a small part of it tunnels out --- this gives rise to a small \emph{echoes} in the escaping wave, which repeat at periodic intervals after the initial, fast, black hole-like ringdown.

\footnotetext{In particular, this is the potential for a Solodukhin wormhole with $\lambda=10^{-10}$ \cite{Bueno:2017hyj,Dimitrov:2020txx}.}

If the effective potential that is felt by the wave looks roughly like a square cavity of length $L$ (as in fig. \ref{fig:echopotential}), one can easily extract the general features of the resulting low-frequency quasinormal mode spectrum  of the compact object. The real part $\omega_R$ of quasinormal modes will be quantized in units of the inverse cavity size.\footnote{This size is measured in an appropriate tortoise coordinate (which corresponds to the travel time of a null geodesic), see section 4.1 in \cite{Cardoso:2016rao}.} The (small) imaginary part $\omega_I$ simply arises from multiplying the tunneling probability $|\mathcal{A}|^2$ of the wave to tunnel through the cavity wall with the frequency with which it hits the outer wall (which is $\omega_R$), so we get the scaling:
\be \label{eq:genomegascaling} \omega = \omega_R + i\, \omega_I, \qquad \omega_R \sim \frac{n}{L}, \qquad \omega_I \sim  |\mathcal{A}|^2\, \omega_R \sim \omega_R^{2\ell+3},\ee
where we used $|\mathcal{A}|^2\sim \omega_R^{2\ell+2}$ \cite{Cardoso:2019rvt} and assumed a wave of angular wavenumber\footnote{In \cite{Cardoso:2019rvt}, $\ell$ is a four-dimensional wavenumber. The same scaling also holds in five dimensions as well \cite{Dimitrov:2020txx}.} $\ell$.

\begin{figure}[ht]\centering
 \includegraphics[width=0.7\textwidth]{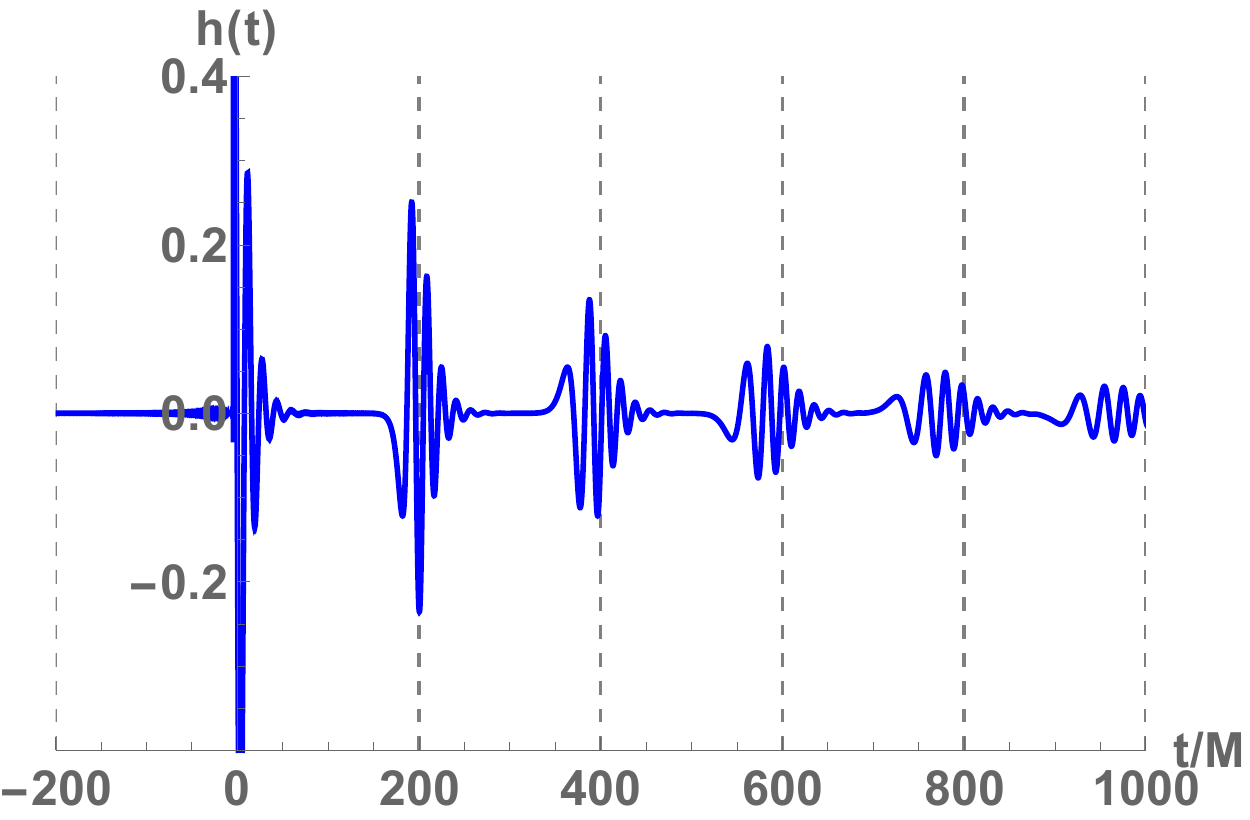}
 \caption[Ringdown echoes]{The (initial) ringdown at $t\sim 0$ in a gravitational wave signal $h(t)$, followed by quasi-periodic and slowly attenuating echoes in the late ringdown signal. The period of the echoes is twice the cavity length, which is taken to be $L = 100M$.\footnotemark}
 \label{fig:echoes}
\end{figure}
\footnotetext{{Figure created from data using the analytic template for echoes in non-spinning ECOs \cite{Testa:2018bzd,GWsite,testathesis}; in particular, the template with $L=d=100M, \mathcal{R}=0.75$, grav. polar $\ell=2$ is used.}}

The echo signal resulting from an effective potential of the form of fig. \ref{fig:echopotential} is also easy to guess: it consists of a periodic signal with period $2L$, which slowly attentuates in amplitude at a timescale $\sim \omega_I^{-1}$ for a given quasinormal mode; see fig. \ref{fig:echoes}. Note that the initial ringdown will always look \emph{as if} it comes from a black hole, since it is (only) sensitive to the light ring of the object (which lies far beyond any structure at horizon scales) \cite{Cardoso:2016rao}. To find the echo spectrum in practice, one can simply decompose the initial ringdown signal in the quasinormal spectrum of the compact object in question, i.e. one finds which modes are excited and how much --- these modes then determine the expected echo waveform in the later ringdown phase \cite{Mark:2017dnq}.

Note that even though the signal coming from the compact object differs only in the (small) echo spectrum in the late ringdown phase, the quasinormal spectra of the compact object and black hole are completely different: the real part of the black hole modes is clearly not quantized in units of $1/L$, and the imaginary part of the compact object's quasinormal frequencies is parametrically smaller than those of the black hole modes (for compact objects in AdS, the imaginary part vanishes completely \cite{Dimitrov:2020txx,Bena:2019azk}). Understanding how a relatively small difference in waveforms can give rise to a completely different quasinormal spectrum was discussed in \cite{Hui:2019aox}.

Clearly, the observation of echoes would be a smoking gun pointing towards the presence of a compact, horizonless object instead of a black hole. There were initial claims of echo observations in the LIGO data \cite{Abedi:2016hgu}, although more detailed data analysis shows this is far from conclusive \cite{Ashton:2016xff}.\footnote{A more complete list of references regarding the discussions on the claims of echo observations can be found in e.g. \cite{Dimitrov:2020txx}, at the start of section 1.} Nevertheless, gravitational wave echoes remain a prime target in gravitational wave searches and modelling of compact object responses. Further comments on the theoretical (un)likeliness of the search for echoes being successful are in section \ref{sec:echoestypical}.


\subsection{Fuzzball echoes \& quasinormal modes}\label{sec:fuzzballechoes}

 A free, probe scalar field in the single-mode $(1,0,n$) superstratum geometry with a long throat ($b^2/a^2\gg1$, see fig \ref{fig:10nsuperstrata}) and $AdS_3\times S^3$ asymptotics was studied in \cite{Bena:2019azk}. In asymptotically AdS spacetime, a natural object to study is the \emph{response function} $\mathcal{R}$, which is the two-point function in the holographically dual boundary CFT of the scalar operator $\mathcal{O}$ dual to the probe scalar field. As a function of time, this AdS response function $\mathcal{R}$ can roughly be seen as the time signal one would see after perturbing the scalar field in the spacetime; in other words, a prototype for the gravitational wave response of the ringdown phase in a merger (see section \ref{sec:GWs} and fig. \ref{fig:GWphases}). (The relation between the AdS response function and the actual signal one would expect in the asymptotically flat spacetime is discussed below.)
 
A novel technique called the \emph{hybrid WKB method} was introduced in order to find the response function in the $(1,0,n)$ superstrata \cite{Bena:2019azk}. Using this technique, it was found that the behaviour of a scalar wave mode in this background depends greatly on its frequency \cite{Bena:2019azk}. At very low frequencies, the behaviour of the scalar wave will be essentially determined by the cap (see fig. \ref{fig:10nsuperstrata}), so that the wave behaves as if it lives in a (redshifted version of) global $AdS_3$ spacetime. At higher frequencies, the wave will begin to explore and feel the effects of the BTZ-like $AdS_2$ throat more. For the response function in position-space, $\mathcal{R}(t)$, this implies that at short- and intermediate time scales, the response function decays just as the corresponding BTZ response function does. However, there is a large time scale $\tau_{\rm echo}$, at which the low-frequency modes conspire to give the response function periodic ``echoes from the cap''. Interestingly, at an intermediate time-scale $\tau_{\rm interm}$, the superstratum response function starts deviating quantitatively from the BTZ one. The two relevant time-scales are set by the parameters of the superstratum solution (see section \ref{sec:SS}):
\be \label{eq:10ntaus} \tau_{\rm echo} \sim \frac{b^2}{a^2}\sim N_1 N_5\, R_y, \quad \tau_{\rm interm} \sim \frac{b}{a}R_y \sim \sqrt{N_1 N_5}\, R_y.\ee
Since the length of the throat in such a superstratum scales as $\sim b^2/a^2$, this scaling of $\tau_{\rm echo}$ precisely conforms with the general expectations of fig. \ref{fig:echoes} and discussed above; the associated (low-frequency) normal modes also conform with the generic scaling expectation (\ref{eq:genomegascaling}). Perhaps more surprising is the intermediate time scale which is \emph{parametrically smaller} than the echo time scale, $\tau_{\rm interm}\ll \tau_{\rm echo}$, at which the response function starts to differ appreciably from the black hole one. Another, possibly related phenomenon in superstrata geometries is related to the geodesic tidal forces, discussed in section \ref{sec:tidalforces}.

It was recently argued \cite{Martinec:2020cml} that when considering a \emph{string} probe (instead of a scalar probe discussed above) in superstrata backgrounds, the resulting echo behaviour can be qualitatively different. For a string, \emph{tidal forces} act to stretch and excite the string, leading to a strong dampening of the string's kinetic energy \cite{Martinec:2020cml}. (See also section \ref{sec:tidalforces} for more discussion on these tidal forces.) The string ``bounces'' back, but does so with less and less kinetic energy at each successive bounce. These bounces of the string may then lead to observable, decaying echoes signals, although these would be different in nature (less sharp, but still periodically spaced) from the scalar echoes discussed above \cite{Martinec:2020cml}.


\subsubsection{Relating AdS and flat space physics}
The response function, $\mathcal{R}(t)$, studied in \cite{Bena:2019azk} and discussed above is an object relevant for a spacetime with $AdS_3\times S^3$ asymptotics, so it is not immediately clear what its relevance is for flat space physics. However, as discussed in section \ref{sec:SS}, the asymptotically flat superstrata must be related to the asymptotically $AdS_3\times S^3$ version by a \emph{decoupling limit} where one takes a dimensionless parameter $\epsilon\rightarrow 0$. If instead one takes the \emph{near-}decoupling limit where $1\gg\epsilon>0$, the general expectation is that the signal at asymptotic flat infinity is essentially given by the response function $\mathcal{R}$ at leading order in $\epsilon$ \cite{Dimitrov:2020txx}. In particular, this means that the echo structure (and timescale $\tau_{\rm echo}$ of (\ref{eq:10ntaus})) of the flat space signal will be the same as that of the $AdS$ analysis discussed above.

At the next, subleading order in the decoupling parameter $\epsilon$, generically the normal modes of $AdS$ will turn into quasinormal modes by receiving a small imaginary part which satisfies the generic scaling (\ref{eq:genomegascaling}) \cite{Avery:2009tu,Chakrabarty:2015foa,Chakrabarty:2019ujg,Dimitrov:2020txx}. This imaginary part can heuristically be seen as arising from the ``leakage'' of the mode out at flat infinity, which was impossible in $AdS$ (as $AdS$ asymptotics behave like a confining box). This generic expectation, including the scaling (\ref{eq:genomegascaling}) of the imaginary part of quasinormal modes, was confirmed in an explicit calculation for a probe scalar field in a single-mode $(2,1,n)$ superstratum geometry using the hybrid WKB method in \cite{Bena:2020yii}.

\subsection{Echoes in typical states?}\label{sec:echoestypical}
The echo structure found in the single-mode $(1,0,n)$ geometries \cite{Bena:2019azk} mentioned above certainly sounds appealing from an observational point of view. However, a valid question is: do we expect to see such echoes in a typical black hole microstate?

First of all, the echoes in the single-mode $(1,0,n)$ geometries come from the low-frequency modes which travel all the way down the throat and explore the ``cap'' microstructure at the bottom of the throat, before escaping back out to infinity. However, if we were to consider a multi-mode $(1,0,n)$ geometry, then the complicated interferences of these modes create extra structure inside the throat \cite{Heidmann:2019xrd}, presumably leading to a qualitatively different echo pattern where the single echoes created by reflection on the bottom of the cap are replaced by a variety of (smaller) echoes which come from (partial) reflections at the various locations in the throat of the mode structures. Although a detailed analysis of the response function in such a multi-mode geometry has not been done, it is reasonable to suspect that the result may not give a very clear, periodic echo signal.

Second, as discussed above in section \ref{sec:MGtypicality}, microstate geometries correspond to a small subspace of coherent microstates, and a typical microstate may have a very different structure. In particular, simple statistical arguments suggest the absence of sharp, observable echoes in any typical microstate \cite{Raju:2018xue,Dimitrov:2020txx}; simply put, when one takes a random (typical) superposition of microstates with sharp echoes, the interference between the echo structures of the different microstates will lead to the echo signal being completely washed out \cite{Dimitrov:2020txx}.



\section{Multipoles \& Tidal Love Numbers}\label{sec:multipolesTLNs}
In this section, I will first give a brief introduction to gravitational multipoles before discussing these multipoles in fuzzballs. I also give a short discussion on tidal Love numbers, which can be understood as an object's multipole deformation response to external tidal stresses.

\subsection{Multipoles}
In electrodynamics, the asymptotic multipole expansion of the potential gives valuable information on the distribution of charges in the interior spacetime.
Since the multipole expansion is an expansion in $1/r$, one may worry that coordinate invariance in general relativity precludes the existence of such a well-defined multipole expansion. Luckily, work by Geroch \cite{Geroch:1970cd}, Hansen \cite{Hansen:1974zz}, and Thorne \cite{Thorne:1980ru} has shown that there exist two families of coordinate-independent 
multipole moments for four-dimensional, asymptotically flat metrics: the mass multipoles $M_{\ell m}$ and the current (or angular momentum) multipoles $S_{\ell m}$. In suitable coordinates \cite{Thorne:1980ru}, these can be read off from the asymptotic expansion of the metric components. When restricting to axisymmetry, so that only $M_{\ell 0}\equiv M_\ell$ and $S_{\ell 0}\equiv S_\ell$ survive, this expansion for $g_{tt}$ and $g_{t\phi}$ is:
\begin{align}
\label{eq:gttmultipoles} g_{tt} &= -1 + \frac{2M}{r} + 2\sum_{\ell\geq2} \frac{1}{r^{\ell+1}} M_\ell P_\ell(\cos\theta) + \text{(harmonics)},\\
\label{eq:gtphimultipoles} g_{t\phi} &= -2r\sin^2\theta \sum_{\ell\geq 1} \frac{S_\ell}{\ell} P_\ell'(\cos\theta) + \text{(harmonics)},
 \end{align}
 where the harmonics are terms that arise at order $r^{-\ell-1}$ that have an angular dependence $P_\ell'(\cos\theta)$ with $\ell'<\ell$,
 see \cite{Thorne:1980ru, Cardoso:2016ryw} for more details. 
 The most familiar multipoles are the mass $M_0\equiv M$ and angular momentum $S_1 \equiv J$. Note also that the mass dipole moment must vanish, $M_1=0$; this can always be achieved by a translation of the origin $r=0$.
 
The Kerr black hole has the multipole structure $M_{2n} = M(-a^2)^n$ and $S_{2n+1} = Ma(-a^2)^n$, with all other multipoles vanishing; this can be compactly written as:
\be M_\ell + i S_\ell = M(ia)^\ell.\ee
It is interesting and perhaps surprising to note that the Kerr-Newman black hole, containing also a charge $Q$, has \emph{the same} multipole moments as Kerr; the charge does not contribute to the multipole moments \cite{Sotiriou:2004ud}!

The multipoles of a given metric in principle capture all of its non-dynamical properties, and it often is insightful to express observable quantities in terms of the multipolar decomposition of the object. For example, the energy $E(\omega)$ of geodesic circular orbits, as a function of the orbital frequency $\omega$, can be related to the multipole moments. Such a geodesic particle will emit gravitational waves at a frequency $2\omega$, and the amount of (logarithmic) energy emitted, $\Delta E(\omega) \equiv -\nu dE/d\nu$, was directly and explicitly related to the multipole moments of the spacetime by Ryan \cite{Ryan:1995zm,Cardoso:2016ryw}.

The lower multipole moments such as the quadrupole moment $M_2$ are often those that are the most interesting (and easy) to observe and analyze (see also section \ref{sec:other} below). For example, $M_2 M_0/S_1^2$ is a simple, dimensionless ratio of multipoles that is exactly equal to 1 for the Kerr black hole. For compact objects, this is not necessarily true, although depending on the model, such objects can still give values for this ratio arbitrarily close to 1 \cite{Cardoso:2016ryw}.

 For a modern review of multipoles in general relativity and  their application to astrophysical observables, see  \cite{Cardoso:2016ryw}, especially Section 5.1 and the Appendix.

 \subsection{Fuzzball multipoles}\label{sec:fuzzballmultipoles}

Four-dimensional, asymptotically flat supersymmetric black holes are static and so have all their multipoles vanishing (except the mass $M=M_0$); so at first thought, they may not seem like an obvious object to its and its microstates' multipolar structure. However, it turns out that dimensionless \emph{ratios} of multipoles have a rich structure, even for these static black holes. Even though formally any ratio of multipoles is undefined for a supersymmetric black hole, it was shown in \cite{Bena:2020see,Bena:2020uup} that using a so-called \emph{indirect} method, these multipole ratios become well-defined quantities characterizing the black hole.
This \emph{indirect} method consists of deforming the black hole in question to a more general, non-extremal charged black hole in four-dimensional maximal supergravity. Then, one calculates the multipole ratio for this deformed black hole. Finally, one simply takes the limit of this multipole ratio as the deformation is turned back off and the original black hole is recovered. Any multipole ratio becomes well-defined (although may be zero or infinite) in this way; note that it is non-trivial that the multipole ratio result obtained in this way is independent of the deformation used.

Since the \emph{indirect} method gives a non-trivial multipole ratio structure to the supersymmetric black hole, the next step is then to compare these ratios with those of some of its microstates in the form of four-dimensional multicentered bubbled geometries (see section \ref{sec:multicenter}). Any such multicentered geometry will generically have all of its multipoles non-vanishing \cite{Bena:2020see,Bena:2020uup,Bianchi:2020bxa,Bianchi:2020miz}. However, in the \emph{scaling limit}, when all centers coincide, the microstate geometry becomes indistinguishable from the black hole and thus shares its zero multipoles. One can then once again consider multipole \emph{ratios}, which are finite away from the scaling point and remain finite and well-defined in the scaling limit --- calculating multipole ratios in this way using scaling multicentered microstate geometries was called the \emph{direct} method in \cite{Bena:2020see,Bena:2020uup}.

Comparing the \emph{indirect} and \emph{direct} methods gives interesting results. For some black hole/microstate geometry pairs, the two methods do not agree well, but for others, they give suprisingly and incredibly close results. It was conjectured in \cite{Bena:2020see,Bena:2020uup} that the methods will agree better when the corresponding black hole has a small so-called \emph{entropy parameter} $\mathcal{H}$, which is the (square of the) dimensionless ratio of the entropy of the black hole over the entropy of a modified black hole with the same electric charges and zero magnetic charges. It would be interesting to explore this conjecture further, for example for the almost-BPS black holes (see section \ref{sec:otherMG}) and their multicentered microstates \cite{almostBPSinprogress}; these black holes have the added advantage of allowing non-zero angular momentum and thus have non-zero multipoles.

Incidentally, the \emph{indirect} method can also be applied to any other black hole, and in particular the Kerr black hole \cite{Bena:2020see,Bena:2020uup}. Many multipoles vanish in the Kerr black hole, but using the \emph{indirect} method, any multipole \emph{ratio} becomes well-defined. For example, consider the simple ratio:
\be \frac{M_2 S_\ell}{M_{\ell+1}S_1} = 1,\ee
which was previously undefined for $\ell$ even (since then $S_\ell=M_{\ell+1}=0$ for Kerr), but is now well-defined after utilizing the \emph{indirect} method. These multipole ratios lead to constraints on models of modifications of the Kerr black hole and may moreover be measurable as relaxation effects of the gravitational waves emitted during black hole mergers \cite{Bena:2020see,Bena:2020uup}.

Even though multicentered microstate geometries have very different electric and magnetic charges from the Kerr black hole, one can nevertheless compare multipoles and multipoles ratios of the multicentered geometries to Kerr, as was done in \cite{Bianchi:2020bxa,Bianchi:2020miz}.\footnote{Note that \cite{Bianchi:2020bxa,Bianchi:2020miz} considers multicentered microstate geometries where the centers are not necessarily on a line and thus a more general version of the axisymmetric multipole formulae (\ref{eq:gttmultipoles})-(\ref{eq:gtphimultipoles}) must be used. Here, I will simply adjust the nomenclature of \cite{Bianchi:2020bxa,Bianchi:2020miz} to the axisymmetric notation using only $M_\ell,S_\ell$ (instead of general $M_{\ell m},S_{\ell m}$) for simplicity of presentation.} After studying a large ensemble of various families of three-center solutions, two main, intriguing results were found:
\begin{itemize}
 \item In \emph{most} of the microstate geometries (90\% or higher of those considered in \cite{Bianchi:2020miz}), multipole ratios such as:
 \be\label{eq:frakMl} \mathfrak{M}_\ell \equiv \left|\frac{M_\ell M_0^{\ell-1}}{S_1^\ell}\right|,\ee
 are \emph{larger} than the corresponding ones in Kerr; for example: $\mathfrak{M}_2 > (\mathfrak{M}_2)^{\text{Kerr}}=1$.
 \item The ratios $\mathfrak{M}_\ell$ (as well as similar ones replacing $M_\ell$ by $S_\ell$ in (\ref{eq:frakMl})) are \emph{monotonically increasing functions} of the intercenter distance, achieving a minimum value at the scaling point where the centers coincide. This was found for \emph{all} of the microstate geometries considered in  \cite{Bianchi:2020miz}.
\end{itemize}
For neither of these phenomena is it known whether they should be expected to hold more generally or not (for example, in multicentered geometries with more than three centers). There are clearly still many interesting questions to explore. One other avenue could be to explore almost-BPS, rotating black holes and their microstates (see section \ref{sec:otherMG}), and to compare their multipolar structure to Kerr \cite{almostBPSinprogress}. Moving away from supersymmetry in this way would be a step in the right direction towards realistic string theory predictions for deviations from the multipolar structure of Kerr.


  \subsection{Tidal Love numbers}\label{sec:TLNs}

The multipole moments discussed above are fundamental but \emph{non-dynamical} properties of a black hole (or compact object). Indeed, they are defined in a setting where the object is alone in spacetime. By contrast, in dynamical processes such as two-body mergers, one would expect an object to deform from the tidal stresses due to the other object. The dynamical rate at which such external tidal stresses deform an object's gravitational multipole moment is called the \emph{tidal Love number} associated to the object for a given multipole. This deformability leaves imprints on the phase of the gravitational waveform (it enters as a fifth-order post-Newtonian correction), especially in the late inspiral orbit phase (see fig. \ref{fig:GWphases}) \cite{Cardoso:2017cfl,Giddings:2019ujs}. In certain cases, the effect from the tidal Love numbers may even be ``magnified'' and affect the gravitational waveform at \emph{leading} order in the mass ratio \cite{Pani:2019cyc}.

Whereas reading off the multipoles of an object simply require a judicious choice of asymptotic coordinate system, calculating Love numbers is a much more daunting task. In principle, given a solution describing a particular black hole or compact object, one must first calculate the general, time-independent linear perturbation to this solution in the full gravity theory at hand. Then, given the general solution that is regular at the interior (or the horizon, if applicable), at each multipole order one takes the ratio of the (polynomially) decaying and growing modes at spatial infinity to find the Love number.

As an example which also serves as a prototype for the actual gravitational Love number calculation \cite{Kol:2011vg}, consider a free, probe scalar field $\Phi$ in a Schwarzschild black hole background in $d$ dimensions. At spatial infinity, the time-independent solution behaves as:
\be \Phi \sim  \left( c_1\, r^\ell + c_2\, r^{-\ell-d+3}\right) Y_{\ell\vec{m}} .\ee
Demanding regularity at the horizon fixes $c_2$ in terms of $c_1$, and the Love number behaves as \cite{Kol:2011vg}:
\be \label{eq:TLN} \lambda_\ell \sim \frac{c_2}{c_1} \sim \tan \left(\pi \frac{\ell}{d-3}\right)   .\ee
(I am foregoing extra factors here which are unimportant to our discussion; in particular, the normalization of the Love numbers in the literature are not always the same.)

In four dimensions, we immediately conclude from (\ref{eq:TLN}) that the Schwarzschild black hole \emph{has all zero (electric\footnote{The correct definition of the magnetic Love numbers was only recently clarified \cite{Pani:2018inf}.}) Love numbers} \cite{Damour:2009vw,Binnington:2009bb,Gurlebeck:2015xpa}. This is actually rather surprising, since the black hole horizon does deform due to stresses \cite{Gurlebeck:2015xpa} , but clearly these deformations must conspire to keep the distribution of mass and thus the multipoles constant. A charged, static black hole in Einstein-Maxwell theory also has vanishing Love numbers \cite{Cardoso:2017cfl}. However, for the rotating Kerr black hole, it is not clear if the Love numbers vanish for all possible gravitational perturbations.\footnote{Various statements can be found in the literature such as ``The tidal Love numbers of black holes in four dimensions vanish''; I wish to emphasize that, to the best of my knowledge, this has \emph{only} been convincingly proven for static black holes in Einstein-Maxwell gravity. It has also been shown that the Love numbers vanish for axisymmetric perturbations of slowly rotating black holes \cite{Landry:2015zfa,Pani:2015hfa}. Recently, it has been argued that more general gravitational perturbations lead to non-zero Love numbers for Kerr \cite{LeTiec:2020spy}, although \cite{Chia:2020yla} claims that the Kerr Love numbers \emph{do} vanish when the correct boundary conditions are used.
Note also that generically, black holes in alternative theories of gravity do not have vanishing Love numbers \cite{Cardoso:2017cfl,Cardoso:2018ptl}. See also section 1 of \cite{Hui:2020xxx} and references therein for an overview. }

The situation is a bit more confusing in five dimensions. There, we see that (\ref{eq:TLN}) implies the Love numbers are either vanishing (for even $\ell$) or infinite (for odd $\ell$). These divergences can be understood as similar to QFT divergences in that they lead to a (classical) RG flow; by separating $\lambda$ into an (infinite) counterterm and a finite part, one can regularize (\ref{eq:TLN}) and extract the finite, observable part of the Love number \cite{Kol:2011vg}. In any case, even for a simple non-rotating black hole in dimensions other than four, the tidal Love numbers do not necessarily vanish.

Compact objects (such as boson stars, gravastars, or wormholes) generically have Love numbers that differ from those of black holes \cite{Cardoso:2017cfl}; in particular, \emph{static} compact objects will typically have non-zero Love numbers. As such, it would be very interesting to calculate the Love numbers of microstate geometries and compare them to their respective black hole Love numbers. However, calculating the full linear perturbation in supergravity to such geometries seems to be a gargantuan task, and has not yet been attempted. Instead, as a prototype calculation, one could consider a free, probe scalar field in the microstate background --- similar to the discussion in \cite{Kol:2011vg} which led to (\ref{eq:TLN}). Such (prototype) scalar Love numbers could then be computed in and compared between black hole and microstate geometries \cite{TLNinprogress}.

 

\section{Geodesics \& Shadows}\label{sec:geodesics}

Since microstate geometries introduce intricate microstructure replacing the black hole horizon, it is natural to wonder how geodesics will behave as they come close to this structure. In particular, the absence of horizons suggests that microstate geometries will not absorb any incident light rays. This could lead to large and measurable differences of microstate geometry shadows compared to black hole ones. Unfortunately, no images of microstate geometries have been simulated yet (to compare with e.g. fig. \ref{fig:EHT}), although some work is in progress \cite{raytracingmicrostatesinprogress,tomlorenzoinprogress}. Also, the framework for imaging five-dimensional black objects was recently introduced \cite{Hertog:2019hfb} (see also \cite{Ahmed:2020dzj,Belhaj:2020okh}), setting the stage for comparisons of microstate geometries and black objects in higher dimensions.

Although images of microstate geometries have yet to be made, there are various aspects of geodesics in microstate geometries that have already been studied in recent years. I will briefly review a few such topics below.

\subsection{Geodesics trapping \& instabilities}

An analysis of geodesics in the single mode $(1,0,n)$ superstrata\footnote{Null geodesics in single-mode $(1,0,n)$ and $(2,1,n)$ superstrata geometries have remarkable integrability properties \cite{Bena:2017upb,Walker:2019ntz}, making it feasible to study their properties (semi-)analytically.}  indicate that these geometries \emph{trap} incoming geodesics having a particular impact parameter \cite{Bianchi:2018kzy,Bianchi:2017sds}. This is to be compared to black holes, which absorb any geodesic with an impact parameter below a certain critical value. One can conjecture that this blackness property of black holes arises as a collective effect whereby each microstate can absorb specific channels of geodesics \cite{Bianchi:2018kzy}.

Trapping of geodesics is a general feature of microstate geometries. For multicentered bubbled microstate geometries, this phenomenon was studied in \cite{Eperon:2016cdd} (see also \cite{Keir:2016azt,Eperon:2017bwq}), where it was shown that stably trapped, zero energy (as seen from infinity) null geodesics exist on the so-called \emph{evanescent ergosurfaces} of these geometries (these are timelike surfaces of infinite redshift) \cite{Gibbons:2013tqa}.\footnote{Relatedly, there is a Penrose process for microstate geometries \cite{Eperon:2017bwq,Bianchi:2019lmi}.}
 This is evidence that microstate geometries generically have \emph{classical, non-linear} instabilities.\footnote{Certain non-supersymmetric microstate geometries, having an actual ergoregion, can be shown to be \emph{linearly} unstable \cite{Cardoso:2005gj}.} It is not completely clear where this instability might drive the evolution of the microstate geometry to, although there are arguments that the instability might trigger a transition to a more typical microstate (geometry) \cite{Marolf:2016nwu} (see also section \ref{sec:MGtypicality}).

 \subsection{Tidal forces}\label{sec:tidalforces}
 One can also consider the \emph{tidal forces} in microstate geometries due to the geodesic deviation of families of nearby geodesics. Rather surprisingly, these tidal forces become Planckian long before geodesics reach the microstructure replacing the horizon (such as the ``cap'' in superstrata, see section \ref{sec:SS}) \cite{Tyukov:2017uig,Bena:2018mpb,Bena:2020iyw,Martinec:2020cml}. This is a generic feature for geometries which look asymptotically like the black hole, but have an interior that is capped off \cite{Bena:2018mpb}.
 
 For superstrata, these large tidal forces are experienced at a throat depth of $r\sim \sqrt{ab}$ \cite{Tyukov:2017uig,Bena:2018mpb} (using single mode $(1,0,n)$ parameters as introduced in section \ref{sec:SS}) which corresponds to a time scale $\tau_{\rm tidal}\sim \sqrt{N_1N_5}\,R_y$. Note that this is the same time scale as the intermediate time scale $\tau_{\rm interm}$ discussed in section \ref{sec:fuzzballechoes} at which the superstratum scalar correlator starts to differ appreciably from the BTZ one. It is interesting to note that a holographic interpretation in the dual CFT of this ``intermediate'' throat depth $r\sim \sqrt{ab}$ or tidal time scale $\tau_{\rm tidal}\sim \sqrt{N_1N_5}\,R_y$ is not known \cite{Bena:2018mpb,Martinec:2020cml}.
 
 Similar results of tidal forces were found for generic multicentered bubbled geometries \cite{Bena:2020iyw}, although one can also fine-tune the parameters in such microstate geometries to push the position at which large tidal forces are experienced deeper down the throat.
 
 Finally, one can also consider probe \emph{strings} in superstrata geometries. The large tidal forces stretch and excite such strings, causing the string to lose (or rather convert) its kinetic energy \cite{Martinec:2020cml}. Interestingly, while the tidal forces experiences are large, the time that the string experiences them is small, and so the net effect of these forces is limited. This will lead to the string being able to ``bounce'' back from the cap (without escaping); after successive bounces it will eventually get trapped deep inside the microstate geometry throat. Such ``tidal trapping'' may be the mechanism at which stringy degrees of freedom can get effectively trapped in horizonless microstate geometries. As mentioned above in section \ref{sec:fuzzballechoes}, the successive bounces of the string may then lead to observable, decaying echoes in gravitational wave signals \cite{Martinec:2020cml}.
 
These tidal effects on probe strings may also have consequences for black hole (microstate) images. In particular, the analysis of \cite{Martinec:2020cml} can be interpreted to indicate that massless particles (such as photons) can become massive at horizon scales due to these stringy, tidal interactions. This may reveal itself in features of images at the EHT that deviate from expectations using null geodesic ray tracing.\footnote{I wish to thank R. Walker for discussions on this point.}

\subsection{Chaotic behaviour}
A closely related subject to that of instabilities of infalling probes is that of chaos and scrambling. A signal of quantum chaos is to take a (normalized) commutator squared of two operators that commute at $t=0$:
\be C(t) \sim \left| \langle \left[ \mathcal{O}_1(0),\mathcal{O}_2(t)\right]^2\rangle\right|,\ee
which also can be related to a certain out-of-time-order correlator (OTOC) of $\mathcal{O}_{1,2}$ for $t$ much larger than the inverse temperature of the system. At early times in a chaotic system, one expects this commutator to grow exponentially in a way governed by the Lyapunov exponent $\lambda_L$, $C(t)\sim e^{\lambda_L t}$. This persists until a time $t\sim \tau_{\rm s}$  called the \emph{scrambling time}, which is when the commutator $C(t)\sim \mathcal{O}(1)$. For $t\gtrsim \tau_{\rm s}$, the behaviour of $C(t)$ is determined by quasinormal mode decay. See \cite{Craps:2020ahu,Polchinski:2015cea} for further details.

This Lyapunov exponent is perhaps the most studied measure of chaos for black holes and their microstates. Black holes have been argued to be \emph{maximally chaotic, fast scramblers} \cite{Maldacena:2015waa}, which means that the black hole Lyapunov exponent would be a universal upper bound on a system at temperature $T$: $\lambda_L \leq 2\pi T$. Note that the scrambling time $\tau_{\rm s}$ for a black hole can be seen as the time necessary for a black hole to thermalize after a perturbation; black holes being fast scramblers means they thermalize very quickly.\footnote{Thermalization in the D1/D5 system from the dual CFT point of view and in the context of fuzzballs was discussed in \cite{Hampton:2019csz}.}

Unstable null geodesics such as circular null orbits close to a black hole are a signal for chaotic behavior in a spacetime.
The Lyapunov exponent $\lambda_L$ can be seen as the inverse of the instability timescale associated with this geodesic motion \cite{Cardoso:2008bp}; it can also be shown to be related to the late-time decay behaviour of quasinormal modes of the geometry.
The Lyapunov exponent was studied by means of unstable null geodesics in the single-mode $(1,0,n)$ superstrata geometry. It was found that the superstrata Lyapunov exponent is generically lower than the black hole value \cite{Bianchi:2020des}; through the connection with late-time dominating quasinormal modes, this may lead to ringdown signals (see section \ref{sec:GWs}) which could discriminate microstate geometries from black holes. OTOCs in superstrata geometries were studied in detail in \cite{Craps:2020ahu}, for example finding that the tidal time scale $\tau_{\rm tidal}$ is much bigger than the scrambling time scale, $\tau_{\rm s}\ll \tau_{\rm tidal}$.

\section{The Future of Fuzzball Observations}\label{sec:other}

\paragraph{Next steps}
Beyond those discussed in this review, there are many other observables accessible within the microstate geometry program that would be interesting for future analysis. For example, resonance effects in the inspiral phase (see section \ref{sec:GWs}) are closely related to tidal effects (see sections \ref{sec:TLNs} and  \ref{sec:tidalforces}) and may lead to effects such as high frequency ``glitches'' in or a phase shift of the gravitational wave signal \cite{Asali:2020wup,Fransen:2020prl}. Also related is the ``tidal heating'' effect \cite{Datta:2019euh,Datta:2019epe,Datta:2020gem,Datta:2020rvo}, which affects the energy loss in the inspiral phase and is especially relevant for EMRIs \cite{Datta:2019euh,Datta:2019epe}.
Another observable is the effect of interactions of microstates with their accretion discs, which may differ appreciably from their black hole counterparts \cite{Macedo:2013jja,Macedo:2013qea}.

Moreover, there is also much more one can do with the analysis of the observables discussed in this paper. For one, the echo analysis discussed in section \ref{sec:fuzzballechoes} was done for the \emph{single-mode} $(1,0,n)$ superstrata; one might expect the sharp echo structure to be changed in a \emph{multi-mode} geometry where the long microstate throat is no longer featureless. Many questions also still remain regarding multipole moments and their ratios (see section \ref{sec:fuzzballmultipoles}), and the study of microstate geometry tidal Love numbers has yet to begin in earnest (see section \ref{sec:TLNs}). There are also many open questions regarding the behaviour of geodesics in microstate geometries, of which perhaps the most obvious and pressing is to produce images and shadows resulting from microstate geometries to compare to the corresponding black hole image (see section \ref{sec:geodesics}).

Another avenue of thought is to use general arguments within the fuzzball paradigm, without making reference to any specific class microstate geometries. This has the advantage of bypassing the limitations of the known microstate geometries (see section \ref{sec:limitations}), but with the downside of being restricted to general plausibility arguments without any concrete calculations. Using such generic arguments, it was reasoned that a novel kind of gravitational wave burst could be observed, associated with the quantum transitions involved in the fuzzball formation process in gravitational collapse \cite{Hertog:2017vod}. Such arguments were also used in \cite{Guo:2017jmi} to discuss the observability of fuzzballs on general grounds.

\paragraph{Experimental sensititivies}

The current studies of observables in microstate geometries are mostly qualitative and not quantitative, since the geometries correspond to non-physical black holes. Nevertheless, it is interesting to briefly note the expected sensitivities of current and future experiments for a few observables discussed above:
\begin{itemize}
 \item Echoes from an ECO with perfect reflectivity (as assumed in the discussion in section \ref{sec:echoesQNMs}) might be detected or ruled out at $5\sigma$ confidence level by LIGO and VIRGO \cite{Testa:2018bzd}.
 \item It is estimated that the future space-based gravitational wave detector eLISA could detect changes in the Kerr quadrupole $M_2$ up to a precision of $\delta M_2/M^3 < 10^{-4}$ \cite{Barausse:2020rsu}.
\item Love numbers and tidal effects thereof would most likely not be measured by ground-based detectors such as LIGO, but could be detected by eLISA. For example, eLISA will be able to put constraints on ECOs with $\epsilon\gtrsim 2/3$ (using the closeness parameter introduced in section \ref{sec:intro}) up to 1\% accuracy for ECO masses around $M/M_{\odot} \sim 10^4-10^6$, and moreover could explore tidal forces of supermassive ECOs with even smaller $\epsilon$ \cite{Cardoso:2017cfl}.
\end{itemize}
Unfortunately, EHT observations are currently not expected to be very sensitive in distinguishing black holes from ECOs, especially when the ECOs are very compact ($\epsilon\ll 1$) \cite{EHT2019e}; this is due to the plasma physics of the accretion disc being largely insensitive to the physics inside the light ring (see section \ref{sec:EHT} and also section \ref{sec:echoesQNMs}). Nevertheless, the theoretical shadows (that is, with an idealized light source) can be shown for other ECOs to be significantly different than those of black holes \cite{Cunha:2015yba}, so it remains an interesting question to understand what distinguishing features fuzzball shadows might have (see section \ref{sec:geodesics}).

\paragraph{Outlook}

Fuzzballs provide the only top-down approach to producing black hole microstates within quantum gravity which replace singularities and horizons with smooth, geometric horizon-scale microstructure. This makes them exciting examples of exotic compact objects (ECOs), of which the gravitational phenomenology can be studied to analyze how they might give rise to observable signals in current and future experiments. The fuzzballs that can be explicitly constructed --- the microstate geometries --- do not correspond to realistic black holes. Nevertheless, studying them can give qualitative, general insights into the physics and observability of the microstructure that we expect to see at black hole horizon scales, and provide a working toy model to estimate for how quantum gravity effects lead to observable deviations from general relativity.

The area of fuzzball and microstate geometry phenomenology is a budding new field where many exciting insights and observations lie ready for the picking.
Despite the complexity of microstate geometries and their limitations (see section \ref{sec:limitations}), the existing analyses have already led to many new and interesting realizations and understandings into the observable physics of microstate geometries.
I truly hope that both string theorists and gravitational phenomenologists will work together in the coming years to further investigate the role of string theory and fuzzballs in predicting and modelling possible deviations from classical general relativity that can be measured at current and future black hole observations.


\section*{Acknowledgments}
I would like to thank the editors of the General Relativity and Gravitation Topical Collection on The Fuzzball Paradigm for the invitation to contribute this review.
I would also like to thank I. Bena, B. Ripperda, D. Turton, B. Vercnocke, A. Virmani, R. Walker, and N. Warner for many insightful and interesting discussions,  as well as suggestions and comments on this paper.
I am especially grateful to B. Vercnocke for introducing me to this research avenue and for many enthousiastic discussions over the years.
I am supported by ERC Advanced Grant 787320 - QBH Structure and ERC Starting Grant 679278 - Emergent-BH.

\appendix

\bibliographystyle{toine}
\bibliography{fuzzballobservations}

\end{document}